\documentclass[a4paper,12pt]{article}

\usepackage{ifpdf}

\newif\ifpdf
\ifx\pdfoutput\undefined
  \pdffalse
\else
  \pdfoutput=1
  \pdftrue
\fi

\RequirePackage{xspace} %
\RequirePackage{subfigure} %
\RequirePackage[centertags]{amsmath} %
\RequirePackage{amssymb}
\RequirePackage{wrapfig} %
\RequirePackage{calc} %
\RequirePackage{ifthen}
\RequirePackage{tabularx} %
\RequirePackage{flafter} %
\RequirePackage{fancyhdr} %

\ifpdf
  \RequirePackage[pdftex]{color}%
  \RequirePackage{colortbl}%
  \RequirePackage{array}%
  \RequirePackage[pdftex]{graphicx}

  \RequirePackage[ pdftex, plainpages = false, pdfpagelabels,
                 pdfpagelayout = useoutlines,
                 bookmarks,
                 breaklinks = true,
                 linktocpage,
                 colorlinks = true,
                 linkcolor = blue,
                 urlcolor  = blue,
                 citecolor = blue,
                 anchorcolor = blue,
                 hyperindex = true,
                 hyperfigures
                 ]{hyperref}

\else
  \RequirePackage{color}
  \RequirePackage{colortbl}
   \RequirePackage{array}
  \RequirePackage[dvips]{graphicx}
  \RequirePackage{hyperref}
  \usepackage{rotating}
\fi

\usepackage{makeidx} % enable indexing
\usepackage{setspace} %enable setting spacing, %e.g. singlespace, onehalfspace, and doublespace
\usepackage{rotating} %enable sidewaysfigure
\usepackage{ecltree}
\usepackage{epic}
\usepackage{supertabular}  % this will allow the table not to break
\usepackage{color}
\usepackage{exscale}
\usepackage{fontenc}
\usepackage{ifthen}
\usepackage{latexsym}
\usepackage{makeidx}
\usepackage{syntonly}
\usepackage{inputenc}
\usepackage{graphicx}
\usepackage{setspace}
\usepackage{caption2}
\usepackage[english]{babel}
\usepackage[square, comma,numbers,sort&compress]{natbib}
\usepackage{hypernat}
\usepackage{boxedminipage}
\usepackage{framed}
\usepackage{longtable}
\usepackage[all]{hypcap}    %included to make the hyperlink go to the top of figure
\usepackage{algorithm2e}
\usepackage{algorithmic}
\usepackage{lscape}
\usepackage{pdflscape}
\usepackage[T1]{fontenc}
\usepackage{microtype}
\usepackage{enumitem}

\newcommand{\note}[1]{\marginpar[left]{\singlespace \tiny #1}}

\renewcommand{\sectionmark}[1]%
      {\markright{\thesection\ #1}} %stops it capitalizing. #1 has value of section name

\renewcommand{\note}[1]{}

\onehalfspace %\doublespace

\DisableLigatures[f]{encoding = *, family = * }

\begin{document}
\begin{center}
{\Large Reply to ``Comment on Sochi's variational method for generalised Newtonian flow'' by
Pritchard and Corson}
\par\end{center}{\Large \par}

\begin{center}
Taha Sochi
\par\end{center}

\begin{center}
{\scriptsize University College London, Department of Physics \& Astronomy, Gower Street, London,
WC1E 6BT \\ Email: t.sochi@ucl.ac.uk.}
\par\end{center}

\vspace{1cm}

\noindent In a recent communication by Dr David Pritchard and Dr Lindsey T. Corson (henceforth PC)
to be published in this Issue of Rheologica Acta, they criticized the use of the variational
principle and the derived variational formulation to obtain flow relations for generalized
Newtonian fluids (GNF) in circular pipes which was proposed by the author in a recent publication
in Rheologica Acta \cite{SochiVariational2013} (hereafter we abbreviate this reference with RA).
The criticism is mainly based on the variational formulation given by Eq. (9) of RA, i.e.

\begin{equation}\label{var}
\frac{\partial}{\partial r}\left(\mu \frac{d\gamma}{dr} \right) = 0
\end{equation}
where $r$ is the tube radius, $\mu$ is the generalized Newtonian viscosity and $\gamma$ is the rate
of strain.

Although we admit that PC have legitimate concerns that should be addressed, in general we either
oppose their views or do not accept the implications and conclusions that they try to reach. In the
present communication we respond in brief to PC criticisms in the following points.

\begin{enumerate}[leftmargin=0cm,itemindent=.6cm,labelwidth=\itemindent,labelsep=0cm,align=left]

%XXXXXXXXXXXXXXXXXXXXXXXXXXXXXXXXXXXXXXXXXXXXXXXXXXXXXXXXX
\item

There are two separate issues: one is the variational principle and the other is the particular
variational formulation (e.g. Euler-Lagrange as in \cite{SochiVariational2013} or Dirichlet as in
\cite{SochiTubeCircSA2015}). The failure of a particular formulation does not mean a failure of the
variational principle on which it is based because the failure of the former may be caused by
mathematical technicalities and not necessarily by the failure of the working principle. Similarly,
the failure of one formulation does not necessarily imply the failure of the other.

%XXXXXXXXXXXXXXXXXXXXXXXXXXXXXXXXXXXXXXXXXXXXXXXXXXXXXXXXX
\item

We believe that, at least for the investigated 1D flow systems, the variational principle, whose
essence is minimizing the total stress, can be established by a mathematical argument based on the
Dirichlet principle \cite{SochiTubeCircSA2015}. We also believe that it can be established
independently by a physical argument based on the linearity (or more precisely proportionality to
the conduit velocity-varying spatial dimension) of the stress function in the investigated 1D flow
systems, i.e. circular pipes and plane slits. Although we do not intend to make any definite
proposal here about the physical foundation of the variational principle, it could be an instance
of energy minimization which proved in the past to be a valid and useful principle in many
mechanical systems.

%XXXXXXXXXXXXXXXXXXXXXXXXXXXXXXXXXXXXXXXXXXXXXXXXXXXXXXXXX
\item

We generally disagree with PC claim that the method is shown to be exact only for power law (PL)
fluids. The method, according to RA, should be exact for non-viscoplastic fluids. Yes, mathematical
difficulties may hinder the analytical progress at some stages in going from $\gamma$ to $v$ (flow
speed) to $Q$ (volumetric flow rate) and hence numerical methods (mainly numerical integration and
numerical solvers like bisection as well as numerical evaluation of complicated analytical
functions like hypergeometric) should be used leading to possible gathering of numerical errors. In
\cite{SochiTubeCircSA2015} the Dirichlet-based variational approach was formulated in a more
general way to provide exact solutions, considering its numerical nature, even for viscoplastic
fluids.

%XXXXXXXXXXXXXXXXXXXXXXXXXXXXXXXXXXXXXXXXXXXXXXXXXXXXXXXXX
\item

In our view, the use of partial derivative, instead of ordinary derivative, in Eq. \ref{var} is a
trivial notational issue as long as we agree on how the equation should be applied. Yes there is a
more important issue about the derivation of this equation which will be discussed in the
forthcoming point about explicit and implicit dependency and its significance on the use of
ordinary and partial derivatives. What is important is that, in our view Eq. \ref{var} as applied
in RA and subsequent publications is correct as will be shown later.

%XXXXXXXXXXXXXXXXXXXXXXXXXXXXXXXXXXXXXXXXXXXXXXXXXXXXXXXXX
\item

The seeming suggestion that our test cases indicate a PL-type approximation of the investigated
instances, even if it can be accepted for the RA fluid models and the given examples, is difficult
to accept for the models investigated in the later publications \cite{SochiSlitPaper2014,
SochiVarNonNewt2014, SochiCarreauCross2015, SochiTubeCircSA2015} where nine different rheologies in
total, each with several examples, have been considered for the flow in circular pipes and plane
slits.

%XXXXXXXXXXXXXXXXXXXXXXXXXXXXXXXXXXXXXXXXXXXXXXXXXXXXXXXXX
\item

To address the technical concerns of PC about the derivation of Eq. \ref{var} as expressed in ``The
mathematical basis of the variational method'' of their Comment, we now re-derive this equation
taking into account PC concerns. We have \footnote{We use $G$ slightly differently from the one
used by PC where our $G$, although still represents the magnitude of the pressure gradient,
contains a dimensionless geometric factor. The purpose of this is to simplify the notation and to
be more general by accommodating different geometries.}

\begin{equation}\label{eqTG1}
\frac{d\tau}{dr}=\frac{d\left(\mu\gamma\right)}{dr}=G
\end{equation}
that is

\begin{equation}\label{eqTG2}
\mu\frac{d\gamma}{dr}+\gamma\frac{d\mu}{dr}=G
\end{equation}
As in RA, we use the following form of the Euler-Lagrange variational equation

\begin{equation}
\frac{d}{dx}\left(f-y'\frac{\partial f}{\partial y'}\right)-\frac{\partial f}{\partial x}=0
\end{equation}
Following Arfken and Weber \cite{ArfkenWBook2005} \S\ 17, the second term is zero. On substituting,
as in RA, we obtain

\begin{equation}\label{mVar}
\frac{d}{dr}\left(\gamma\frac{d\mu}{dr}+\mu\frac{d\gamma}{dr}-\mu\frac{d\gamma}{dr}\right)=0
\end{equation}
i.e.

\begin{equation}\label{var1}
\frac{d}{dr}\left(\gamma\frac{d\mu}{dr}\right)=0
\end{equation}
This is the first from of the variational principle. Alternatively, we substitute from Eq.
\ref{eqTG2} into Eq. \ref{mVar} to obtain

\begin{equation}
\frac{d}{dr}\left(G-\mu\frac{d\gamma}{dr}\right)=0
\end{equation}
which reduces to

\begin{equation}\label{var2}
\frac{d}{dr}\left(\mu\frac{d\gamma}{dr}\right)=0
\end{equation}
because $G$ is independent of $r$ (in fact for the considered cases of circular pipes and plane
slits with rigid walls and uniform cross sections $G$ is constant because the pressure gradient in
the flow direction is constant). Eq. \ref{var2} is the second form of the variational principle
which is the same as Eq. \ref{var} apart from the notational difference about partial and ordinary
derivative. Despite their apparent difference, the two forms (Eq. \ref{var1} and Eq. \ref{var2})
are essentially the same because the two terms, $\mu\frac{d\gamma}{dr}$ and
$\gamma\frac{d\mu}{dr}$, add to a constant, so a variation in one term will be compensated by a
variation by the same amount in the other term as can be seen from Eq. \ref{eqTG2}. This means that
the variational process will be fully represented by any one of the two terms as the essence of
this process is present in each one of them independently. Adding to a constant also means that the
system dynamics is fully represented by each one of the two terms and hence exact solutions can be
obtained from each one of these forms although each one contains only a single term of the full
``dynamical'' form (Eq. \ref{eqTG2}). As there is nothing in this derivation to indicate a
dependency on a particular type of fluid, the formulation should be valid for all types of GNF not
only PL fluids. This will address the issue of limited validity of Eq. \ref{var2} to PL fluids
which is claimed by PC. This issue will be examined further next.

%XXXXXXXXXXXXXXXXXXXXXXXXXXXXXXXXXXXXXXXXXXXXXXXXXXXXXXXXX
\item

We come now to the issue related to PC argument leading to Eq. (11) of their Comment and the
conclusion that Eq. \ref{var2} holds exact only for PL fluids. We express our objection to PC
argument where they apply the ``dynamical'' and variational formulations simultaneously by imposing
the variational condition on the ``dynamical'' form. Hence, the ``if and only if'' part in the
argument should be re-examined. To be more specific about this point we say that the two
variational forms are based on absorbing the dynamics of the absent term into the existing term.
Hence, when we substitute from the variational form into the ``dynamical'' form the other term in
the ``dynamical'' form will vanish automatically not because it is zero but because it is
represented by the existing term in the substituted variational form. So, as soon as PC substitute
from the second variational form into their Eq. (6), the first term in this equation will vanish
and hence their Eq. (7) will reduce to

\begin{equation}
\frac{d}{dr}(0)=0
\end{equation}
which is trivial and hence no condition can be derived from it. This is not because
$\gamma\frac{d\mu}{dr}$ is actually zero but because its dynamic effect is absorbed by the other
term in the substituted variational form. To clarify this point in a more formal way we integrate
the two variational forms as well as the ``dynamical'' form in a generic fashion to obtain

\begin{equation}\label{i1}
\int\gamma d\mu=Ar+B
\end{equation}

\begin{equation}\label{i2}
\int\mu d\gamma=Cr+D
\end{equation}

\begin{equation}\label{i3}
\int\gamma d\mu+\int\mu d\gamma=Gr+F
\end{equation}
where $A,\, B,\, C,\, D$ and $F$ are constants. To obtain the results of the first form we
substitute into the ``dynamical'' form (Eq. \ref{i3}) from Eq. \ref{i2} to absorb the effect of the
second term and obtain

\begin{equation}
\int\gamma d\mu+Cr+D=Gr+F\,\,\,\,\,\,\,\,\,\Rightarrow\,\,\,\,\,\,\,\,\,\,\,\int\gamma d\mu=\left(G-C\right)r+(F-D)
\end{equation}
Similarly, we obtain the results of the second form by substituting into the ``dynamical'' form
(Eq. \ref{i3}) from Eq. \ref{i1} to absorb the effect of the first term and obtain

\begin{equation}
Ar+B+\int\mu d\gamma=Gr+F\,\,\,\,\,\,\,\,\,\Rightarrow\,\,\,\,\,\,\,\,\,\,\,\int\mu d\gamma=\left(G-A\right)r+(F-B)
\end{equation}
So the three forms give the same type of correlation but with different sets of constants which
should be adjusted properly if the rules are followed and the boundary conditions are applied
correctly.

Let us now assume that our objection in the last paragraph is invalid and hence Eq. (11) of PC
Comment holds true. We now ask PC to explain why we can get exact solutions for PL from a single
term, i.e. $\mu\frac{d\gamma}{dr}$? Does this mean that the other term, i.e. $\gamma
\frac{d\mu}{dr}$, in the ``dynamical'' equation has zero contribution? Certainly not! It can be
shown that the same exact solution for PL can be obtained from applying the first variational form,
i.e. Eq. \ref{var1}, which employs this term, as it was obtained in RA from applying the second
variational form, Eq. \ref{var2}, which employs the other term. Hence the exact analytical solution
can be obtained not only from the ``dynamical'' form which contains both terms but also from each
one of the variational forms where each one involves only a single term. The reason, in our view,
is that the system dynamics is fully represented and contained in each one of the two terms due to
the fact that the two terms add to a constant and hence exact analytical solutions for PL can be
obtained from each one of the variational forms. For the same reason, exact solutions should also
be obtained for non-PL fluids because this property is not specific to PL. In fact we need an
explanation not only for getting exact solutions for PL fluids from the variational forms but also
for getting ``approximations'' for non-PL fluids since either term looks as important as the other.
Neglecting the effect of one term, if it really happened as PC seem to suggest, should lead to
wrong solutions that cannot be considered even as bad approximations.

We now provide another argument to endorse the general applicability of Eq. \ref{var2}, as well as
Eq. \ref{var1}, to all GNF and not only to PL fluids. The stress, according to the ``dynamical''
formulation, is a function of the conduit geometry and pressure drop but not of the fluid rheology,
as can be seen for example from Eq. (3) of PC Comment where $\tau$ is defined in terms of $p$, $z$
and $r$; none of which is a rheological attribute. Hence for a given conduit and pressure drop the
stress function is the same for all types of GNF. Therefore if we combine: (a) the variational
formulation is based on minimizing this universal stress function with no involvement of the fluid
rheology, and (b) the formulation leads to exact solutions for the PL fluids; then the logical
conclusion is that the variational formulation should also lead to exact solutions for the other
types of GNF. In other terms; the objective of the variational approach is to optimize the stress
function, which is universal and independent of the fluid rheology. If the obtained universal
stress function through this optimization process proved to be correct for one type of rheology,
i.e. PL, by producing an exact solution for this type, then it should also be correct and produce
exact solutions for the other types of rheology due to its universal nature and independence of
rheology.

%XXXXXXXXXXXXXXXXXXXXXXXXXXXXXXXXXXXXXXXXXXXXXXXXXXXXXXXXX
\item

With regard to PC argument about ($\tau_t = \tau_w - \tau_c$), their criticism is based on judging
one approach by the spirit of another approach as they seem to assume that while we are applying
the variational method the stress function is well known whereas if we have to use the variational
approach sensibly we should assume that the stress function is unknown and we, through functional
optimization, try to obtain it. PC seem to assume that the justification of the variational
approach should be based on our real ignorance of the stress function and therefore we are told
that we know this already by the force balance argument and hence the variational approach is
pointless, whereas in reality it is based on assuming that the stress function is subject to a
variational process as if it is unknown; otherwise for all the investigated cases not only the
stress function but also the final solutions are well known from other methods. The linearity as
proposed by ($\tau_t = \tau_w - \tau_c$) when we assume a variational process will be the product
of this process rather than a given fact although we know it in advance from other methods. The
variational principle is based on assuming that the shape of the stress function is optimal, and
hence by applying the variational rules we obtain useful mathematical formulations from which we
can obtain all the parameters of the flow field including the velocity profile. The purpose of the
variational approach then is to get equations like \ref{var1} and \ref{var2} through functional
optimization which is motivated by the search for the stress function as if it is unknown. Hence we
can see nothing in PC argument that undermines the rationale of this method and the assumptions on
which it is based. The value of the variational approach then comes from obtaining alternative
formulations (e.g. Eq. \ref{var1} and Eq. \ref{var2}) for rheological and fluid dynamics problems
that could be easier to use plus possible future extensions to more problems in which the stress
function is unknown or more difficult to obtain from other methods. The value of the variational
approach will be discussed further later.

As for the seeming suggestion of PC of a ($0=0$) ``absurdity'' as part of this argument, first we
do not regard this as an absurdity since ($0=0$) is a legitimate mathematical statement. We also
disagree with PC about the implication of this on the stress minimization, as discussed above.
Moreover, to make a productive use of mathematics it should be harnessed by beneficial objectives
to avoid reaching such trivial results. It is not unusual to come across such an experience where
one route leads to the solution while another, which is also legitimate, leads to a dead end by
producing trivial results like ($0=0$). We can show many examples like this where ($0=0$) can be
produced from correct equations and legitimate operations. For example, we can take Eq. \ref{eqTG2}
and differentiate both sides to obtain

\begin{equation}
\frac{d}{dr} \left( \mu\frac{d\gamma}{dr}+\gamma\frac{d\mu}{dr} \right)=0
\end{equation}
We then substitute from Eq. \ref{eqTG2}, that is

\begin{equation}
\frac{dG}{dr} = 0
\end{equation}
to obtain ($0=0$). There is nothing wrong or illegitimate in this sort of manipulation, but it is
not a productive way of exercising mathematics. Anyway, if PC argument is valid and the variational
formulation is really based on an absurdity or triviality like this then it is hard to believe that
we can obtain from such a formulation exact solutions for the PL fluids and ``approximations'' for
other types of fluid.

%XXXXXXXXXXXXXXXXXXXXXXXXXXXXXXXXXXXXXXXXXXXXXXXXXXXXXXXXX
\item

Regarding PC contemplation about the relation between the ``dynamical'' and variational
formulations, in our view, supported by the derivations and discussions in the previous points, the
two formulations are equivalent in the sense that they both fully represent the system dynamics and
hence they should provide mathematically equivalent solutions although they may be in different
forms. Apart from the theoretical value of having a correct variational formulation, there are
practical advantages as well since the variational form offers an alternative to the ``dynamical''
form where the former could be simpler and hence easier to manage and manipulate analytically and
discretize numerically. Anyway, the variational form should be a legitimate alternative even
without any practical benefits. The theoretical and practical values of the variational approach
can become more significant if it can be extended and generalized to more complex rheologies and
geometries beyond the relatively simple cases of GNF flow through pipes and slits.

%XXXXXXXXXXXXXXXXXXXXXXXXXXXXXXXXXXXXXXXXXXXXXXXXXXXXXXXXX
\item

Now let us assume that all the arguments presented by PC in their Comment are sound, including all
the direct and hinted implications, then the maximum conclusion that can be drawn is that Eq.
\ref{var2} is not general and the variational principle as presented in RA is not well-established.
In our view, this outcome will have little impact on the development of the variational approach
because (A) The variational principle can be established by other ways. In fact we do not even need
to establish the variational principle as a hard fact to legitimize the variational approach; all
we need is the possibility of being a valid principle which cannot be denied as long as there is no
hard evidence against it. The obtained solutions require validation to verify their correctness as
with other methods. (B) Eq. \ref{var2} remains exact for some types of rheology where even PC
accept this fact. (C) The potential limitation of this formulation will not affect the other
variational formulation, i.e. Dirichlet-based formulation \cite{SochiTubeCircSA2015}, which is not
based on Eq. \ref{var2} or the Euler-Lagrange principle. In \cite{SochiTubeCircSA2015} it was
demonstrated that the Dirichlet-based approach produces exact solutions considering its numerical
nature. (D) There is still a possibility, which seems to be accepted by PC, that this
``PL-restricted'' formulation can be a good approximation to non-PL fluids. From our personal
experience, as demonstrated in \cite{SochiVariational2013, SochiSlitPaper2014, SochiVarNonNewt2014,
SochiCarreauCross2015, SochiTubeCircSA2015}, we can assert that this ``approximation'' is a good
one.

\end{enumerate}

\vspace{0.3cm}

Finally, the total or partial success of the variational principle and its formulations in the
investigated 1D flow systems could be coincidental or fundamental; the latter is still a
possibility. As far as this objective is concerned, very little, if any, will change by PC Comment.
In our publications related to this issue, the variational principle and method were advocated not
as hard facts but as possibilities where we demonstrated that by assuming the validity of the
variational principle and employing the proposed variational formulations, definitely or
tentatively correct solutions and logical trends can be obtained and hence, we argued, it is
possible that the flow systems, or certain types of which, are subject to a variational balance. To
change the outcome, we need a definite challenge to the proposed principle and formulations where
it can be shown with certainty that the variational approach breaks down. Although this is
possible, we cannot see such a challenge in the attempt of PC despite its elaboration. However, we
agree with PC about the necessity for more research about this issue.

%XXXXXXXXXXXXXXXXXXXXXXXXXXXXXXXXXXXXXXXXXXXXXXXXXXXXXXXXX

\bibliographystyle{unsrt}

\end{document}